\documentclass[preprint,showpacs,preprintnumbers,amsmath,amssymb]{revtex4}
\usepackage{epsf,amsmath,amssymb,verbatim,color,multirow,pifont}
\usepackage{graphicx}
\newcommand{\av}[1]{\langle #1 \rangle_{\lambda}}
\begin{document}

\title{Priority queues with bursty arrivals of incoming tasks}
\author{N.~\surname{Masuda}$^1$}
\author{J. S.~\surname{Kim}$^2$}
\author{B.~\surname{Kahng}$^2$}
\affiliation{{$^1$ Graduate School of Information Science and Technology,
The University of Tokyo, 7-3-1 Hongo, Bunkyo, Tokyo 113-8656, Japan}\\
{$^2$ Department of Physics and Astronomy, Seoul National
University, Seoul 151-747, Korea}}
\date{REceived 6 September 2008}

\begin{abstract}
Recently increased accessibility of large-scale digital records
enables one to monitor human activities such as the interevent
time distributions between two consecutive visits to a web portal
by a single user, two consecutive emails sent out by a user, two
consecutive library loans made by a single individual, etc.
Interestingly, those distributions exhibit a universal behavior,
$D(\tau)\sim \tau^{-\delta}$, where $\tau$ is the interevent time, and
$\delta \simeq 1$ or $3/2$. The universal behaviors have been
modeled via the waiting-time distribution of a task in the queue
operating based on priority; the waiting time follows a power law
distribution $P_{\rm w}(\tau)\sim \tau^{-\alpha}$ with either $\alpha=1$ or
$3/2$ depending on the detail of queuing dynamics. In these models,
the number of incoming tasks in a unit time interval has been assumed
to follow a Poisson-type distribution. For an email system, however,
the number of emails delivered to a mail box in a unit time we measured
follows a powerlaw distribution with general
exponent $\gamma$. For this case, we
obtain analytically the exponent $\alpha$, which is not necessarily $1$
or $3/2$ and takes nonuniversal values depending on
$\gamma$. We develop the generating function formalism to obtain
the exponent $\alpha$, which is distinct
from the continuous time approximation
used in the previous studies.
\end{abstract}


\maketitle

\section{Introduction}
In the digital era, human activities can be easily monitored and
quantified by analyzing digital records such as the dates of sending
or replying to emails, and financial transactions.  Interestingly,
human activities generate emerging patterns: the interevent time
distribution of human activities follows a power law, and its exponent
is either 1 or 3/2 in many cases~\cite{Barabasi,Vazquez05,Vazquez06,goh,dezso}.
Such a bursty nature of human dynamics has been understood to be
a consequence
of queuing processes driven by human decision making. Barab\'asi
introduced a queuing model operating in the priority-based
protocol~\cite{Barabasi}. At each time step, a task arrives at such a
queue and is assigned a priority $x_i$ chosen randomly
from a distribution
$\rho(x)$. Then, with probability $p$, the task with the highest
priority is selected for execution and removed from the list. With
probability $1-p$, a task is randomly selected irrespective of its
priority and is executed. This model was successful in analytically
reproducing the empirical result~\cite{Barabasi,Vazquez05,Vazquez06}:
the waiting time of a task in the queue before being executed,
which is denoted by $\tau$, follows a power-law distribution
$P_{\rm w}(\tau)\sim
\tau^{-1}$. The result is independent of distribution
$\rho(x)$. The power law $P_{\rm w}(\tau)\sim \tau^{-3/2}$ is
reproduced by allowing the queue length to vary in
time~\cite{Barabasi,Vazquez06}.

\begin{figure}[!h]
\centerline{\epsfxsize=15cm \epsfbox{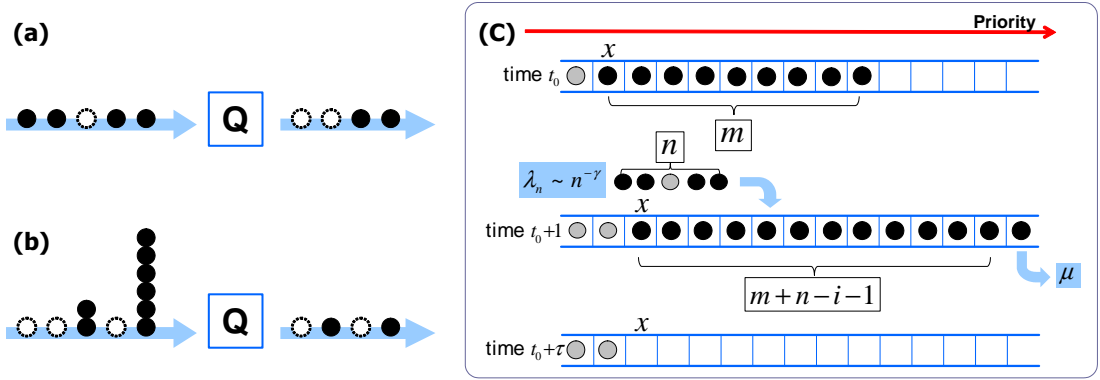}}
\caption{(Color online) Schematic representation
of queueing protocols. (a) A queue system proposed by Grinstein-Linsker, in
which at most one task (filled circle) arrives in the system per time step. (b)
The queue system we consider in this paper, in which input tasks (filled circles) 
can be bursty. (c) Operation of the queue system shown in (b):
At time $t_0$, there are $m$ tasks (black circles) with priority $\ge x$ in the queue.
At time $t_0+1$, $n$ tasks (black and gray circles) arrive in the queue
with probability $\lambda_n$. Among them, $n-i$ tasks (black, not gray, circles)
have priority $\ge x$. This event occurs with probability
$\binom{n}{n-i}\left(1-x\right)^{n-i}x^{i}$, where $0\le i \le n$. The
task with the largest priority is executed with probability $\mu$. No
task is executed with probability $1-\mu$. At time
$t_0+\tau$, the queue does not contain any tasks with priority $\ge x$
for the first time.}\label{queueing}
\end{figure}

To analyze both fixed-length and flexible-length queues,
the Barab\'asi's model
was extended as follows. In each time step, a task arrives with probability
$\lambda$, and the task with the highest priority in the queue list is
executed with probability $\mu$.
Operation of this queue system is schematically shown in
Fig.~\ref{queueing}(a).
Since the dynamics of the queue is
stochastic if $0<\lambda<1$ or $0<\mu<1$,
the queue length generally changes
in time. This model is a type of
the M/G/1 queuing system with a priority selection rule proposed in
the seminal work of Cobham in 1954~\cite{cobham}.
This model was analytically studied
recently.
The waiting-time distribution of a task in the queue changes depending
on $\lambda$ and $\mu$. (i) When
$\lambda=\mu=1$, the number of tasks in the queue is fixed,
and the waiting time of tasks
obeys $P_{\rm w}(\tau)\sim \tau^{-2}$~\cite{Gabrielli07}.
(ii) When
$\lambda=\mu < 1$, $P_{\rm w}(\tau)\sim \tau^{-3/2}$~\cite{Grinstein}.
(iii) When
$\lambda < \mu < 1$, $P_{\rm w}(\tau)\sim \tau^{-3/2}e^{-\tau/\tau_0}$
for $\tau \ll \tau_0$ and $P_{\rm w}(\tau)\sim
\tau^{-5/2}e^{-\tau/\tau_0}$ for $\tau \gg \tau_0$, where the
characteristic time scales as
$\tau_0=1/(\sqrt{\mu}-\sqrt{\lambda})^2$~\cite{Grinstein}.
(iv) When $\mu < \lambda < 1$, tasks with priority
$x < (\lambda-\mu)/\lambda$
wait in the queue forever without being executed.
Tasks with priority $x \ge (\lambda-\mu)/\lambda$
are executed with the waiting time $\tau$
following $P_{\rm w}(\tau)\sim \tau^{-3/2}$~\cite{Grinstein}.

\begin{figure}[!h]
\centerline{\epsfxsize=10cm \epsfbox{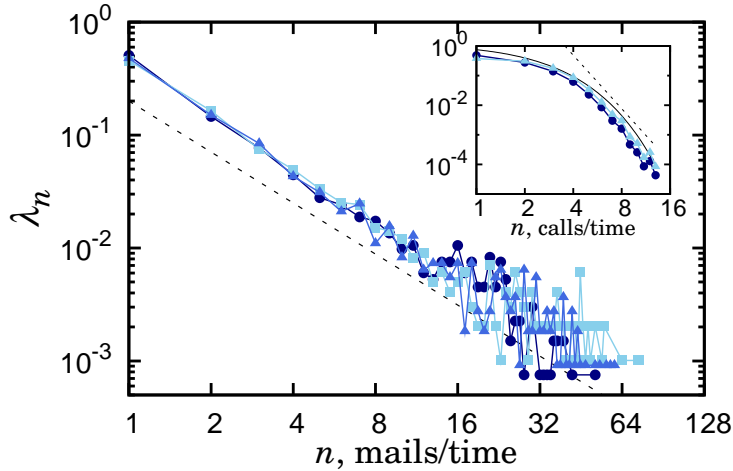}}
\caption{(Color online) Distributions of the number of
incoming tasks. The main panel shows the number of
tasks delivered to an email box of an anonymous
user per unit time~\cite{eckmann}, which follows a power-law distribution with slope $-1.5$.
Different lines correspond to different bin sizes, namely, 500 ($\bigcirc$), 800 ($\triangle$), 
and 1000 ($\square$) seconds.
Note that the slope $-1.5$ is not universal. It depends on users and
can be as small as $-3$. We chose a user with the largest dataset. (Inset) The number of
wireless phone calls arriving at a cell station in 10 seconds for the peak time (i.e., 12:00-20:00)
($\triangle$) and for the entire day
($\bigcirc$). Both data fit well to the Poisson distribution 
(black solid line),
which decays even faster than the power law with exponent $-6$ (dotted line).}\label{fig_numdel}
\end{figure}

Previous studies focused on the case in which incoming tasks are
independent of each other and delivered to the queue at a constant
rate. Thus, the number of incoming tasks in a unit time follows
the Poisson distribution. This is the case observed in, for
example, the number of requests for wireless phone calls arriving
at a cell station in a unit time (see inset of
Fig.~\ref{fig_numdel}). However, we observe that the number of
emails received by a single user in a unit time is heterogeneous
and follows a power-law distribution (see Fig.~\ref{fig_numdel}).
Time intervals between consecutive tasks arriving at a server computer
\cite{Dewes03,Paxson95,Kleban03,Harder06}
and between a user's hypertext markup language (HTML)
requests \cite{dezso}, which are closely related to the number of
incoming tasks per unit time, also show similar patterns.
The origin of such non-uniform numbers of incoming tasks is not
known yet, but may be consequences of multiple correspondences
with multiple people or self-similar patterns in the number of
data packets arriving at a given router~\cite{self-similar}.
Such bursty arrivals of tasks may significantly change the behavior of
priority queue systems. For example, a more skewed
distribution of the number of incoming tasks per unit time may result in
a more skewed waiting-time distribution of a task $P_{\rm w}(\tau)$, as
briefly suggested in \cite{Kleban03}. In this paper, we
study the waiting-time distribution of a task in the queue
for the case of heterogeneous numbers of incoming tasks.
We find that the universal
power-law exponent $\alpha=3/2$
for $P_{\rm w}(\tau)\sim \tau^{-\alpha}$
occurs as a limited case and obtain other values of
$\alpha$ depending
on the power-law exponent
of the distribution of the number of incoming tasks.

\section{Model}
We study the queue model defined as follows:
in each discrete time step, $n$ tasks are delivered to
the queue, where $n$ is distributed according to a power law
$\lambda_n=\lambda n^{-\gamma}/\zeta(\gamma)$ ($n > 0$),
$\lambda_0=1-\lambda$, where $0\le \lambda\le 1$ and
$\zeta(\gamma)\equiv \sum_{n^{\prime}=1}^{\infty} n^{\prime
-\gamma}$ is the Riemann $\zeta$ function. Each task is assigned a
priority $x$ uniformly distributed on [0,1]. At the same time,
the task with the highest priority in a queue is executed
with probability $\mu$ ($0\le \mu\le 1$). Operation of this queue
system is schematically depicted in Figs.~\ref{queueing}(b) and 
\ref{queueing}(c).
The queue length is unbounded so that the queue accommodates
all incoming tasks. This model generalizes the model introduced
by Grinstein and Linsker (GL)~\cite{Grinstein},
which corresponds to $\lambda_{0}=1-\lambda$,
$\lambda_{1}=\lambda$, and $\lambda_{n}=0$ for $n\ge 2$ in our model.

We will obtain the waiting-time distribution $P_{\rm w}(\tau)$ for a task
in the queue. To this end, we start with
the probability that there are $m$ tasks with priority
larger than or equal to
$x$ in the queue at time $t$, which is denoted by $Q_x(m,t)$.
We denote the queue-length distribution in the steady
state by ${\tilde Q}_x(m)=\lim_{t\to\infty}Q_x(m,t)$. Note that the steady state exists
only under a certain condition, as discussed later.  We define
$G_x(m,\tau)$ to be the probability that a given task with
priority $x$ arriving in the queue at time $t=t_0$ is
executed at time $t=t_0+\tau$.
When the task
arrives in the steady state, there are already $m$
tasks in the queue with priority larger than or equal to $x$,
where $m$ is distributed according to ${\tilde Q}_x(m)$.
All of these $m$ tasks are
executed before the given task is executed.
Then, the waiting-time distribution is obtained via
the following formula \cite{Grinstein}:
\begin{equation}
P_{\rm w}(\tau)=\sum_{m=0}^{\infty} \int_0^1 dx {\tilde Q}_x(m)G_x(m,\tau),
\label{eq:P_w(tau) def}
\end{equation}
where $G_x(m,\tau)$ is equivalent to
the first passage probability that a random walker
starting from position $m>0$ arrives at the origin at time
$\tau$ for the first time.
For a constant rate of incoming tasks,
${\tilde Q}_x(m)$, $G_x(m,\tau)$, and $P_{\rm w}(\tau)$ can be
obtained explicitly \cite{Grinstein}.
However, due to the complexity of our problem,
we obtain them implicitly in terms of the generating functions. We define
the generating function
\begin{equation}
{\mathcal P}_{\rm w}(s) \equiv
\sum_{\tau=1}^{\infty} P_{\rm w}(\tau)s^{\tau},
\end{equation}
where $0< s < 1$.
Then,
\begin{equation}
{\mathcal P}_{\rm w}(s)=\sum_{m=0}^{\infty}\int_0^{1}dx {\tilde Q}_x(m){\mathcal G}_x(m,s),
\label{eq:P_w(s) def}
\end{equation}
where
${\mathcal G}_x(m,s)\equiv \sum_{\tau}G_x(m,\tau)s^{\tau}$.
Because the number of tasks in the queue decreases at most one per
unit time, we obtain
\begin{equation}
G_x(m,t)=\sum_{\tau}G_x(m-1,t-\tau)f_x(\tau),
\label{eq:G_x m m-1}
\end{equation}
where $f_x(t)\equiv G_x(1,t)$.
Equation~\eqref{eq:G_x m m-1}
is expressed in terms of generating functions as
\begin{equation}
{\mathcal G}_x(m,s)={\mathcal G}_x(m-1,s){\mathcal F}_x(s),
\label{eq:calG_x m m-1}
\end{equation}
where ${\mathcal F}_x(s)\equiv \sum_{t=1}^{\infty}f_x(t)s^t$.
Applying Eq.~\eqref{eq:calG_x m m-1}
repeatedly, we obtain
\begin{equation}
{\mathcal G}_x(m,s)={\mathcal F}_x^m(s).
\end{equation}
Then, Eq.~\eqref{eq:P_w(s) def} is written as
\begin{equation}
{\mathcal P}_{\rm w}(s)=\sum_{m=0}^{\infty}\int_0^{1}dx {\tilde Q}_x(m){\mathcal F}_x^m(s)
=\int_0^1 dx {\mathcal {\tilde Q}}_x\left({\mathcal F}_x\left(s
\right)\right),
\label{eq:P_w(s) def final}
\end{equation}
where ${\mathcal {\tilde Q}}_x(z)\equiv \sum_{m=0}{\tilde Q}_x(m)z^m$.

\begin{figure}[!h]
\centerline{\epsfxsize=17cm \epsfbox{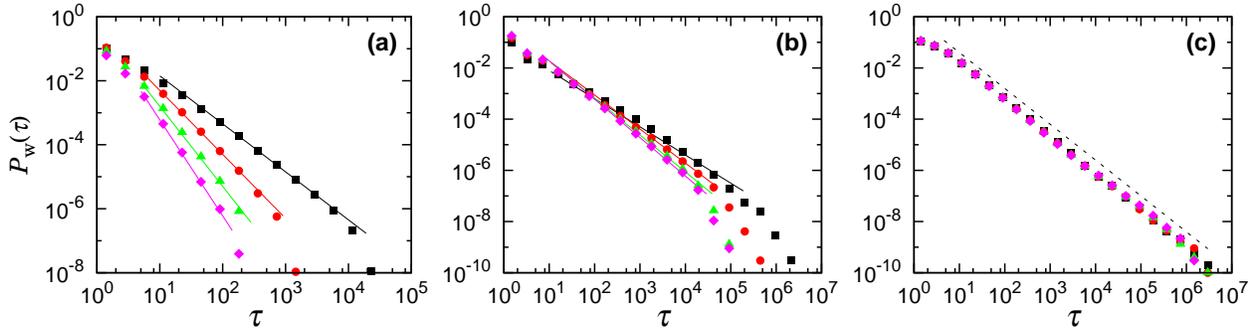}}
\caption{(Color online) The waiting-time distribution $P_{\rm w}(t)$.
(a) The case $\av{n} < \mu$.  Given $\lambda=0.3$ and $\mu=1.0$, shown are numerically obtained $P_{\rm w}(\tau)$ for $\gamma=$ 2.5 ($\square$), 3.0 ($\bigcirc$), 3.5 ($\triangle$), and 4.0 ($\Diamond$), yielding to $\av{n}\approx 0.58$, $0.41$, $0.36$, and $0.33$, respectively.  Solid lines indicate
$P_{\rm w}(\tau)\sim \tau^{-(\gamma -1)}$.  (b) The case $\av{n} \ge \mu$ with $2 < \gamma \le 3$. Given $\lambda=0.5$ and $\mu=0.5$, shown are numerically obtained $P_{\rm w}(\tau)$ for $\gamma=$2.1 ($\square$), 2.5 ($\bigcirc$), 2.8 ($\triangle$), and 3.0 ($\Diamond$), yielding $\av{n}\approx 3.39$, $0.97$, $0.75$, and $0.68$, respectively. Solid lines indicate $P_{\rm w}(\tau)\sim \tau^{-(2\gamma-3)/(\gamma-1)}$. (c) The case $\av{n} > \mu$ with $\gamma > 3$.  Given $\lambda=0.5$ and $\mu=0.3$, shown are numerically obtained $P_{\rm w}(\tau)$ for $\gamma=$ 3.3 ($\square$), 3.8 ($\bigcirc$), 4.0 ($\triangle$), and 4.5 ($\Diamond$), yielding $\av{n}\approx 0.62$,$0.57$, $0.56$, and $0.53$, respectively. The dotted line is a guideline with slope $-1.4$, close to the theoretical value $-1.5$.}\label{fig2}
\end{figure}

Once we derive ${\mathcal {\tilde Q}}_x(z)$ and ${\mathcal F}_x(s)$
explicitly, we obtain the waiting-time distribution of a task in the
queue, namely, $P_{\rm w}(\tau)$. We will show that
the waiting time exhibits a power-law behavior $P_{\rm w}(\tau)\sim
\tau^{-\alpha}$, where the values of $\alpha$ are shown in Table~\ref{table1}.
The analytic solutions are confirmed numerically in Fig.~\ref{fig2}.
Using our generating function formalism, we can also
reproduce the results derived in Ref.~\cite{Grinstein}, as shown in
Appendix~\ref{sec:GL derivation}.

\begin{table}[!h]
\caption{Power-law exponent $\alpha$ of the waiting-time distribution
$P_{\rm w}(\tau)\sim {\tau}^{-\alpha}$.}
\label{table1}
\begin{tabular*}{\hsize}{@{\extracolsep{\fill}}lcr}
~~~~~& $\left<n\right>_{\lambda}<\mu$ &
$\left<n\right>_{\lambda}\ge\mu$~~~~~ \cr
\hline
~~~~~$2<\gamma\le 3$ & $\gamma-1$ &$\frac{2\gamma-3}{\gamma-1}$~~~~~~~ \cr
~~~~~~~~~$\gamma>3$ &  $\gamma-1$ & $\frac{3}{2}$~~~~~~~~~~~\cr
\hline
\end{tabular*}
\end{table}

\section{The queue-length distribution}
In this section, we calculate the queue-length distribution in the
steady state by using the generating function ${\mathcal {\tilde Q}}_x(z)$.
The master equation for $Q_x(m,t)$ is given by
\begin{eqnarray}
Q_x(m,t+1)\hspace{-6pt}&=&\hspace{-6pt}\mu
\sum^{\infty}_{j=0}\lambda_j x^j Q_x(m+1,t)\label{eq:Qm general 1}\\
\hspace{-6pt}&+&\hspace{-6pt} \sum^m_{i=0}\left(1-\mu\right)\sum^{\infty}_{j=i}
\lambda_j \binom{j}{i}\left(1-x\right)^i x^{j-i}Q_x(m-i,t)\label{eq:Qm
general 2}\\
\hspace{-6pt}&+&\hspace{-6pt} \sum^m_{i=0}\mu \sum^{\infty}_{j=i+1}\lambda_j\binom{j}{i+1}\left(1-x\right)^{i+1}
x^{j-i-1}Q_x(m-i,t)\label{eq:Qm general 3}\\
\hspace{-6pt}&\equiv& \hspace{-6pt}\sum^{m}_{i=-1} p_{m-i\to m}Q_x(m-i,t),\quad (m\ge 1).
\label{eq:Qm general 4}
\end{eqnarray}
In the above equation, the three terms in the right-hand side (RHS)
correspond
to different types of events that occur in a unit time.
The first term \eqref{eq:Qm general 1} represents
the case in which $j$
($j=0,1,\cdots$) tasks arrive in the queue with probability $\lambda_j$,
the priorities of all $j$ tasks are smaller than $x$, and
one task is executed with probability $\mu$.
The second term \eqref{eq:Qm general 2} represents the case in which
$j$ ($j=0,1,\cdots$) tasks arrive in the queue with probability
$\lambda_j$, $i$ tasks
out of the $j$ tasks have priorities larger than or equal to $x$, and
no task is executed with probability $1-\mu$.
The third term \eqref{eq:Qm general 3} represents the case in which
$j$ ($j=0,1,\ldots$) tasks arrive in the queue with probability
$\lambda_j$, $i+1$ tasks out of the $j$ tasks have priorities larger
than or equal to $x$, and one task is executed with probability $\mu$.
For later discussion, we denote by $p_{m-i\to m}$ in Eq.~\eqref{eq:Qm general
  4} the transition probability of
the random walk from position $m-i$ to position $m$ in a unit time.
The master equation at the boundary is given by
\begin{eqnarray}
Q_x(0,t+1)\hspace{-6pt}&=&\hspace{-6pt} \mu \sum^{\infty}_{j=0}\lambda_j x^j Q_x(1,t)\nonumber \\
\hspace{-6pt}&+&\hspace{-6pt}\left[\left(1-\mu\right)\sum^{\infty}_{j=0}
\lambda_j x^j + \mu\sum^{\infty}_{j=1}\lambda_j j \left(1-x\right)
x^{j-1} + \mu\sum^{\infty}_{j=0}\lambda_j x^j \right]Q_x(0,t)\nonumber\\
\hspace{-6pt}&\equiv& \hspace{-6pt} p_{1\to 0} Q_x(1,t) + p_{0\to 0}Q_x(0,t).
\label{eq:Qm}
\end{eqnarray}

Based on Eqs.~\eqref{eq:Qm general 1}-\eqref{eq:Qm},
we calculate the generating function
${\mathcal {\tilde Q}}_x(z)$ for the steady-state queue-length distribution
${\tilde Q}_x(m)\equiv \lim_{t\to \infty}{Q}_x(m,t)$. Specifically,
the generating function of Eq.~\eqref{eq:Qm general 1} is equal to
$\mu \Lambda(x) \left[{{\mathcal {\tilde Q}}_x(z)-{\tilde Q}_x(0)}
\right]/{z}$ in
the steady state,
where $\Lambda(z)\equiv\sum^{\infty}_{j=0}\lambda_j z^j$.
The generating function of Eq.~\eqref{eq:Qm general 2} is equal to
$(1-\mu){\mathcal {\tilde Q}}_x(z)\Lambda\left[\left(1-x
\right)z+x\right]$. The generating
function of Eq.~\eqref{eq:Qm general 3} is equal to
$\mu {\mathcal {\tilde Q}}_x(z) \left\{\Lambda\left[\left(1-x\right)z+x\right]-\Lambda\left(x\right)\right\}$.
The generating function of ${\tilde Q}_x(0)$ in Eq.~\eqref{eq:Qm} is
equal to
$\mu \Lambda(x) {\tilde Q}_x(0)$. Combining all these terms,
we obtain
\begin{equation}
{\mathcal {\tilde Q}}_x(z)=\frac{\mu {\tilde Q}_x(0)(z-1)\Lambda(x)}{z-(\mu+z-\mu z)\Lambda\left[
\left(1-x\right)z+x\right]}.
\label{eq:G-final_first}
\end{equation}
To eliminate ${\tilde Q}_x(0)$ from Eq.~\eqref{eq:G-final_first},
we exploit the condition ${\mathcal {\tilde Q}}_x(1)=1$. However,
both the denominator and the numerator of Eq.~\eqref{eq:G-final_first}
converge to zero as $z\to 1$. Thus, we apply the L'Hospital rule
to Eq.~\eqref{eq:G-final_first} to derive
\begin{equation}
{\tilde Q}_x(0)=\left[\mu-
\left(1-x\right)\langle n \rangle_{\lambda}\right]/\left(\mu
 \Lambda\left(x\right)\right),
\label{eq:Q_x(0) stationary}
\end{equation}
where
$\av{n} \equiv \sum_{n=0}^{\infty} n\lambda_n$.
Plugging
Eq.~\eqref{eq:Q_x(0) stationary} into Eq.~\eqref{eq:G-final_first}
yields
\begin{equation}
{\mathcal {\tilde Q}}_x(z)=\frac{\left[\mu-\av{n}\left(1-x\right)\right]
(z-1)}{z-(\mu+z-\mu z)\Lambda\left[
\left(1-x\right)z+x\right]}.
\label{eq:calQ_x(z)}
\end{equation}
For the steady state to exist,
the incoming rate of the task with larger than or equal to $x$
(i.e., $\av{n}(1-x)$) must be smaller than the
execution rate $\mu$ \cite{Barabasi,Grinstein};
$A_1 \equiv \mu - \av{n}(1-x) > 0$ is required. In addition, $\av{n}$ must be finite, which is equivalent to the condition $\gamma>2$.

The mean queue length denoted by $\left<m(x)\right>_{\tilde Q}$ is derived
as
\begin{eqnarray}
\left<m(x)\right>_{\tilde Q} &=&
\left.\frac{\partial {\mathcal{\tilde Q}}_x(z)}{\partial z}\right|_{z=1}
\nonumber\\
&=& \frac{2(1-\mu)\av{n}(1-x) +
(\av{n^2}-\av{n})(1-x)^2}{2A_1}.
\label{eq:m(x)}
\end{eqnarray}
Equation~\eqref{eq:m(x)} implies that
$\left<m(x)\right>_{\tilde Q}$ diverges
when $\av{n^2}$ does, that is, when $\gamma\le 3$.
When $\gamma >3$, the queue length is finite for $x=0$
if and only if $\mu>\av{n}$ and diverges as
$1/\left(\mu-\av{n}\right)$ as $\av{n}$
approaches $\mu$ from below, which extends the results in
\cite{Grinstein}. As $x\to 0$ and $\av{n}\to \mu$,
$\left<m(x)\right>_{\tilde Q}$ diverges as
$1/x$, which is also consistent with the previous
result~\cite{Grinstein}.

To calculate the asymptotic behavior of the steady-state queue-length
distribution $\tilde{Q}_x(m)$, we assume $\mu > \av{n}(1-x)$
and $\gamma > 2$, for which the steady state exists.
When $2<\gamma \le 3$, $\Lambda(z)$ is expanded
near $z\to 1$ as follows \cite{Tauberian}:
\begin{equation}
\Lambda(z) = 1-\av{n}(1-z)
+c_{\gamma}(1-z)^{\gamma-1} + o\left(\left(1-z\right)^{\gamma-1}\right),
\label{eq:Lambda 2-3}
\end{equation}
where $c_{\gamma}$ is a constant.
Inserting Eq.~\eqref{eq:Lambda 2-3}
into Eq.~\eqref{eq:calQ_x(z)} leads to
\begin{equation}
{\mathcal {\tilde Q}}_x(z)=1 - \frac{c_{\gamma}(1-x)^{\gamma-1}(1-z)^{\gamma-2}}{A_1}
+o\left(\left(1-z\right)^{\gamma-2}\right).
\label{eq:calQ_x(z) 2-3}
\end{equation}
For $3<\gamma\le 4$, we obtain
\begin{equation}
\Lambda(z)=1-\av{n}(1-z)+\frac{\av{n^2}-\av{n}}{2}(1-z)^2 -c_{\gamma}(1-z)^{\gamma-1}
+o\left(\left(1-z\right)^{\gamma-1}\right),
\label{eq:Lambda 3-4}
\end{equation}
which leads to
\begin{equation}
{\mathcal {\tilde Q}}_x(z)=1 + \left<m\left(x\right)\right>_{\tilde Q}(z-1)+
\frac{c_{\gamma}(1-x)^{\gamma-1}}{A_1}(1-z)^{\gamma-2}+o\left(\left(1-z\right)^{\gamma-2}\right).
\label{eq:calQ_x(z) 3-4}
\end{equation}
Similar expansions hold true for $\gamma>4$.
By applying the Tauberian theorem \cite{Tauberian}
to Eqs.~\eqref{eq:calQ_x(z) 2-3} and \eqref{eq:calQ_x(z) 3-4}, we obtain
\begin{equation}
\tilde{Q}_x(m)\sim \frac{1}{m^{\gamma-1}} \quad (m\to\infty)
\label{eq:Q-tail}
\end{equation}
for $\gamma >2$. Equation~\eqref{eq:Q-tail} is consistent with the result
under the first-in-first-out (FIFO) protocol~\cite{hglee}.
This is because, when $\mu \gg \av{n}(1-x)$, tasks are executed upon
its arrival in the steady state so that the priority-based protocol
can be regarded as the FIFO-based one.

\section{First-passage probability}
In this section, we derive ${\mathcal F}_x(s)=
\sum_{t=1}^{\infty}f_x(t)s^t= \sum_{t=1}^{\infty}
G_x(1,t)s^t$.  Recall
that $G_x(m,t)$ is the probability that a given task with priority
$x$ is executed at time $t$ after its arrival, provided that there
are $m$ tasks in the queue with priority larger than or equal to $x$
when this task arrives. This quantity can be interpreted as the first
passage probability that a random walker on a half line
starts from position
$m$ and arrives at the origin at time $t$ for the first time. The
probability that the random walker moves from $i$ to $j$
in a unit time is given by $p_{i\to j}$
[see Eq.~\eqref{eq:Qm general 4}].

The generator of the one-step transition of
the random walk before reaching the origin is represented by
\begin{equation}
{\mathcal P}(z)\equiv\sum^{\infty}_{i=-1} p_{m\to m+i} z^i
=\left(1-\mu+\frac{\mu}{z} \right)\Lambda\left[\left(1-x\right)z+x \right].
\label{eq:generator}
\end{equation}
Note that the RHS of Eq.~\eqref{eq:generator} is independent of
$m$ because the transition probability is homogeneous in space.

The amount of a single jump that
the random walker makes to the right is unbounded, because it is equal to
the number of incoming tasks with priority larger than or equal to
$x$. However,
the amount of a jump to the left is at most one,
which yields a useful relation,
\begin{equation}
{\mathcal G}_x(i,s)={{\mathcal F}_x(s)}^i.
\label{eq:calG_x composite}
\end{equation}

Using Eqs.\eqref{eq:generator}, \eqref{eq:calG_x composite},
and the recursion relation~\cite{Deboer94,Masuda05},
\begin{equation}
f_x(t)=p_{1\to 0}+p_{1\to 1}G_x(1,t-1)+p_{1\to 2}G_x(2,t-1)+\cdots,
\end{equation}
we obtain the following
self-consistent equation for the generating function:
\begin{eqnarray}
{\mathcal F}_x(s)&=&s\sum_{i=0}^{\infty}p_{1\to i}{\mathcal G}_x(i,s)\nonumber\\
&=&s\sum_{i=0}^{\infty} p_{1\to i} {{\mathcal F}_x(s)}^i\nonumber\\
&=&
s {\mathcal F}_x(s){\mathcal P}\left({\mathcal F}_x(s)\right)\nonumber \\
&=&s\left[\left(1-\mu\right){\mathcal F}_x\left(s\right)+\mu \right]
\Lambda\left[\left(1-x\right){\mathcal F}_x\left(s\right)+x\right].
\label{eq:calF_x(s) self}
\end{eqnarray}
The first $s$ in the RHS comes from the unit time spent by
a single transition starting from $m=1$.
After this transition, the generating function of the number of tasks
with priority larger than or equal to $x$ in the queue is $z {\mathcal P}(z)$. Since
each such task incurs an execution time distributed according to $\{f_x(t)\}$,
we replace $z$ of $z {\mathcal P}(z)$ by ${\mathcal F}_x(s)$ to obtain
Eq.~\eqref{eq:calF_x(s) self}.

\begin{figure}[!h]
\centerline{\epsfxsize=8cm \epsfbox{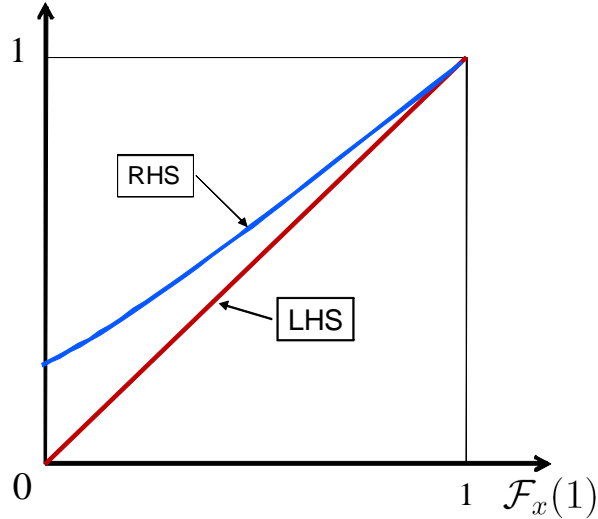}}
\caption{(Color online) Schematic representation of the LHS and the RHS of Eq.~\eqref{eq:fff} as functions of ${\mathcal F}_x(1)$.}
\label{fig33}
\end{figure}

We evaluate ${\mathcal F}_x(s)$ in the limit $s\to 1$ using
Eq.~\eqref{eq:calF_x(s) self}. To guarantee that
the task with priority $x$ is eventually executed,
${\mathcal F}_x(s=1)=1$ has to be satisfied.
To check if this condition is fulfilled, we put $s=1$
in Eq.~\eqref{eq:calF_x(s) self} to obtain
\begin{equation}
{\mathcal F}_x(1)=s\left[\left(1-\mu\right){\mathcal F}_x\left(1\right)+\mu \right]
\Lambda\left[\left(1-x\right){\mathcal F}_x\left(1\right)+x\right].
\label{eq:fff}
\end{equation}
The left-hand side (LHS) and the RHS of Eq.~\eqref{eq:fff} are
plotted in Fig.~\ref{fig33} as functions of
${\mathcal F}_x(1)$,
where ${\mathcal F}_x(1)$ is regarded as a variable for the sake of
this analysis.
Note that the RHS of Eq.~\eqref{eq:fff} is positive
at ${\mathcal F}_x(1)=0$.
Figure~\ref{fig33} implies that
Eq.~\eqref{eq:fff} has the unique solution
${\mathcal F}_x(1)=1$ if and only if the slope of the RHS of
Eq.~\eqref{eq:fff} at
${\mathcal F}_x(1)=1$ is less than or equal to unity, that is,
\begin{equation}
\frac{\partial}
{\partial {\mathcal F}_x(1)}
\left[\left(1-\mu\right){\mathcal F}_x(1)+\mu \right]
\Lambda\left[\left(1-x\right){\mathcal F}_x\left(1 \right)+x\right]
\big|_{{\mathcal F}_x(1)=1} \le 1.
\label{condition_15}
\end{equation}
Equation~\eqref{condition_15} is equivalent to $A_1 \ge 0$,
which is what we already assumed.

In the following, we obtain the solution of the self-consistent
equation~\eqref{eq:calF_x(s) self} by
assuming that $f_x(t)$ follows a power law.

\noindent
{\sf Case (i): $\mu > \av{n}$.} In this case, $A_1 >0$ holds for all $x$.
When $2<\gamma\le 3$, combining Eqs.~\eqref{eq:Lambda 2-3} and \eqref{eq:calF_x(s) self}
yields
\begin{equation}
{\mathcal F}_x(s)=1+\frac{1}{A_1}(s-1)+
\frac{c_{\gamma}(1-x)^{\gamma-1}}{A_1^{\gamma}}(1-s)^{\gamma-1}
+o\left(\left(1-s\right)^{\gamma-1}\right).
\label{eq:calF_x(s) 2-3}
\end{equation}
When $3<\gamma\le 4$, combining
Eqs.~\eqref{eq:Lambda 3-4} and \eqref{eq:calF_x(s) self} yields
\begin{eqnarray}
{\mathcal F}_x(s)=1 + \frac{1}{A_1}(s-1)+ \frac{A_1-A_1^2+A_2}{A_1^3}(s-1)^2
-\frac{c_{\gamma}(1-x)^{\gamma-1}}{A_1^{\gamma}}(1-s)^{\gamma-1} + o\left(\left(1-s\right)^{\gamma-1}\right),
\label{eq:chi-gamma-3-4}
\end{eqnarray}
where $A_2\equiv (\av{n^2}-\av{n})(1-x)^2/2+
(1-\mu)\av{n}(1-x)>0$. Note that the coefficient of
$(1-s)^2$ is positive.
In a similar manner, we can show for $\gamma > 4$ that the leading singular
term of ${\mathcal F}_x(s)$ is equal to $(-1)^{\lceil \gamma \rceil-1}
c_{\gamma}(1-x)^{\gamma-1}(1-s)^{\gamma-1}/A_1^{\gamma}$,
where $\lceil \gamma \rceil = \min\{i ; i\ge \gamma, i\in {\mathbf Z}\}$.
Thus, we obtain $f_x(t)\sim t^{-\beta}$ with $\beta=\gamma$
for $\gamma>2$ using the Tauberian theorem \cite{Tauberian}.

\noindent
{\sf Case (ii): $\mu = \av{n}$.} Because $A_1=0$ for $x=0$,
we cannot apply the results obtained for case (i).
For example, $1/A_1=1/\left(\av{n}x\right)$
in the coefficient of $(s-1)$ in Eq.~\eqref{eq:calF_x(s) 2-3}
diverges as $x\to 0$, implying that the exponent $\beta$
is smaller than 2 near $x=0$.  Actually the long-time behavior
of $f_x(t)$ is dominated by the tasks whose priority is near
$x=0$~\cite{Grinstein}. Thus, we assume
\begin{equation}
{\mathcal F}_x(s)=1-c_{\beta}(1-s)^{\beta-1}+o((1-s)^{\beta-1})
\label{eq:calF_x(s) 1-2}
\end{equation}
with $1 < \beta \le 2$.

When $2<\gamma \le 3$, the RHS of Eq.~\eqref{eq:calF_x(s) self} is written as
\begin{eqnarray}
 &=& s\left[(1-\mu){\mathcal F}_x(s)+\mu\right]\Lambda\left[(1-x){\mathcal F}_x(s)+x\right]\nonumber \\
&=& s{\mathcal F}_x(s)+\mu s \left[1-{\mathcal F}_x(s)\right]+\nonumber \\
&& s\left\{\av{n}\left(1-x\right)\left[{\mathcal F}_x(s)-1\right]+ c_{\gamma}(1-x)^{\gamma-1}\left[1-{\mathcal F}_x(s)\right]^{(\gamma-1)}+\cdots\right\}
\end{eqnarray}
Plugging Eq.~\eqref{eq:calF_x(s) 1-2} into the LHS and RHS of Eq.~\eqref{eq:calF_x(s) self}
leads to
\begin{eqnarray}
&&(1-s)\left[1-c_{\beta}(1-s)^{\beta-1}+\cdots\right]\nonumber\\
&=&\av{n}xc_{\beta}(1-s)^{\beta-1}
+c_{\gamma}(1-x)^{\gamma-1}c_{\beta}^{\gamma-1}(1-s)^{(\beta-1)(\gamma-1)}+\cdots.
\label{eq:near0-gamma-2-3-a}
\end{eqnarray}
If $\av{n}x\gg (1-s)^{(\gamma-2)/(\gamma-1)}$, the first term of the RHS of
Eq.~\eqref{eq:near0-gamma-2-3-a} is much larger than the second term
as $s\to 1$ so that $\beta=2$ and $c_{\beta}=1/\left(\av{n}x\right)$.
Conversely, if $\av{n}x \ll (1-s)^{(\gamma-2)/(\gamma-1)}$, the second term
dominates the first term so that $\beta= 1 + 1/(\gamma-1)$
and $c_{\beta} = c_{\gamma}^{-1/(\gamma-1)}/(1-x)\simeq
c_{\gamma}^{-1/(\gamma-1)}$.

When $3<\gamma< 4$, as in the case of $2 < \gamma \le 3$, Eqs.~\eqref{eq:Lambda 3-4},
\eqref{eq:calF_x(s) self}, and \eqref{eq:calF_x(s) 1-2}, with an appropriate assumption
of $1 < \beta \le 2$, yield
\begin{eqnarray}
(1-s)+o(1-s)&=&\av{n}xc_{\beta}\left(1-s\right)^{\beta-1}
+ A_2 c_{\beta}^2(1-s)^{2(\beta-1)}\nonumber \\
&-&c_{\gamma}(1-x)^{\gamma-1}c_{\beta}^{\gamma-1}
(1-s)^{(\beta-1)(\gamma-1)}+\cdots.
\label{eq:near0-gamma-3-4-a}
\end{eqnarray}
If $\av{n}x\gg \sqrt{A_2(1-s)}$, the first term
in the RHS of Eq.~\eqref{eq:near0-gamma-3-4-a} is much
larger than the second term. Then
$\beta=2$ and $c_{\beta}=1/\left(\av{n}x\right)$.
Conversely, if $\av{n}x\ll
\sqrt{A_2(1-s)}$, the second term is much larger than the first term
so that $\beta=3/2$ and $c_{\beta}=1/\sqrt{A_2}$.
The third term is always much smaller than the second term as $s\to 1$.

\begin{figure}[!h]
\centerline{\epsfxsize=15cm \epsfbox{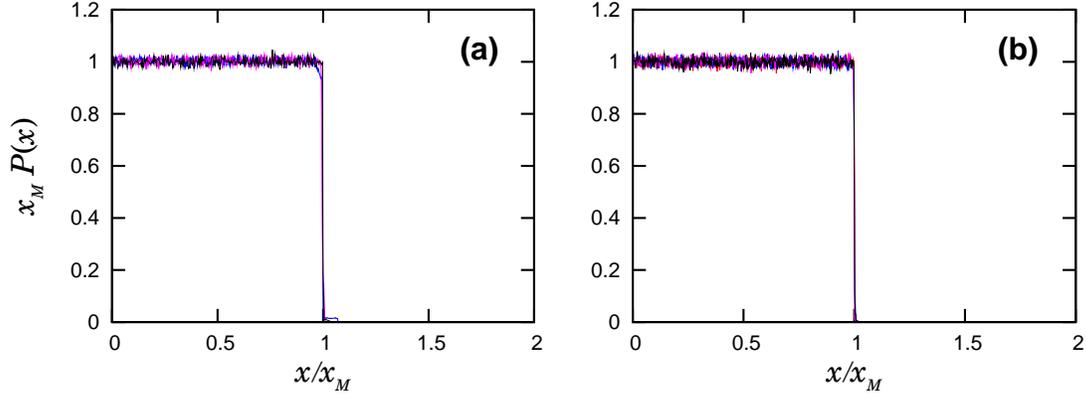}}
\caption{(Color online) Distributions of the priority of tasks in the queue
in the steady state. Distributions of $x$ for different sets
of $\av{n}$ and $\mu$ are shown.
Only the tasks
with priority $x > x_M
\equiv(\av{n}-\mu)/\av{n}$
are executed.
(a) $\lambda=0.5$ and $\mu=0.5$. We set $\gamma=2.1$, 2.5, 2.8, and 3.0,
which yield $x_M=0.85, 0.48, 0.33$, and $0.26$, respectively.
(b) $\lambda=0.5$ and $\mu=0.3$. We set $\gamma=3.3$, 3.8, 4.0, and 4.5,
which yield $0.52$, $0.47$, $0.46$, and $0.43$, respectively.
In each panel,
the four plots almost collapse onto one.}\label{fig:pri_dist}
\end{figure}

\noindent
{\sf Case (iii): $\mu < \av{n}$.} The task in the queue accumulates at rate
$\av{n}-\mu$. In this case, only the tasks with priority
$x> x_M\equiv \left(\av{n}-\mu\right)/\av{n}$ are executed, and the analysis
can be ascribed to case (ii) \cite{Grinstein}. Distributions of the
priority
of tasks in the queue in the steady state are shown
in Fig.~\ref{fig:pri_dist} for some values of $\av{n}$ and $\mu$.

\section{The waiting-time distribution}
Using Eqs.~\eqref{eq:P_w(s) def final}, \eqref{eq:calQ_x(z)}, and ${\mathcal F}_x(s)$ we obtained for the three cases, we calculate
the waiting-time distribution as follows:

\noindent
{\sf Case (i): $\mu > \av{n}$.}
The leading singular term of ${\mathcal {\tilde Q}}_x({\mathcal F}_x(s))$ is
equal to $(-1)^{\lceil \gamma \rceil}
c_{\gamma}(1-x)^{\gamma-1}(1-s)^{\gamma-2}/A_1^{\gamma-1}$.
Then, we obtain
\begin{equation}
{\mathcal P}_{\rm w}(s)\sim (1-s)^{\gamma-2},
\end{equation}
which yields $P_{\rm w}(\tau)\sim {\tau}^{-(\gamma-1)}$
for $\gamma>2$.

\noindent
{\sf Case (ii): $\mu=\av{n}$.} In this case,
we use Eq.~\eqref{eq:calF_x(s) 1-2} with values of
$\beta$ and $c_{\beta}$ depending on $\gamma$ and $x$.

For $2 < \gamma < 3$, we obtain
\begin{eqnarray}
{\mathcal P}_{\rm w}(s)
\hspace{-6pt} &\simeq&\hspace{-6pt}
\int^{(1-s)^{\frac{\gamma-2}{\gamma-1}}}_0 dx
\av{n}x c_{\gamma}^{-\frac{1}{\gamma-1}}
\left(1-s\right)^{\frac{1}{\gamma-1}-1}+\int^1_{(1-s)^{\frac{\gamma-2}{\gamma-1}}}dx+\cdots \nonumber \\
\hspace{-6pt}&=&\hspace{-6pt} 1 + \left(\frac{\av{n}c_{\gamma}^{-\frac{1}{\gamma-1}}}{2}
-1 \right) (1-s)^{\frac{\left(\gamma-2\right)}{(\gamma-1)}}+\cdots.
\label{eq:near0-gamma-2-3-b}
\end{eqnarray}
Therefore, $P_{\rm w}(\tau)\sim \tau^{-(2\gamma-3)/(\gamma-1)}$.

For $3 < \gamma < 4$, we obtain
\begin{eqnarray}
{\mathcal P}_{\rm w}(s) \hspace{-6pt}&\simeq&\hspace{-6pt}
\int_0^{\frac{\sqrt{A_2(1-s)}}{\av{n}}} dx \frac{\av{n}x}
{\sqrt{A_2(1-s)}}+\int^1_{\frac{\sqrt{A_2(1-s)}}{\av{n}}}dx+\cdots \nonumber\\
\hspace{-6pt}&=&\hspace{-6pt} 1-\frac{\sqrt{A_2(1-s)}}{2\av{n}}+\cdots.
\label{eq:near0-gamma-3-4-b}
\end{eqnarray}
Therefore, $P_{\rm w}(\tau)\sim \tau^{-3/2}$.
Similar calculations yield $P_{\rm w}(\tau)\sim \tau^{-3/2}$ for $\gamma>4$.

\noindent
{\sf Case (iii): $\mu < \av{n}$.} Since the analysis can be ascribed to
case (ii), we obtain $P_{\rm w}(\tau)\sim \tau^{-(2\gamma-3)/(\gamma-1)}$
for $2< \gamma\le 3$ and $P_{\rm w}(\tau)\sim \tau^{-3/2}$ for $\gamma>3$.

\section{Discussion and Summary}
The analytic results are summarized in Table~\ref{table1} and
confirmed numerically in Fig.~\ref{fig2}.  The power-law behavior of
the waiting-time distribution $P_{\rm w}(\tau)\sim \tau^{-\alpha}$ can
be diverse in that $\alpha$ can take general values, rather than $\alpha=1$
or $3/2$.  Consistent with this, the intertransaction time of a stock
broker obeys the power-law distribution with $\alpha\approx 1.3$ with
an exponential cutoff \cite{Vazquez06}.

Our results are compatible
with those derived from the continuous time
approximation~\cite{Grinstein} and the fractional
derivative~\cite{klafter}.
The generating function approach that we have developed
can be useful for
studying further problems. For example, we show
in the Appendix that our approach considered
in the limit $\lambda$, $\mu$ $\to 0$ reproduces the
results for the GL model~\cite{Grinstein}.
Furthermore, GL as well as we are successful in deriving the exponential
cutoff for $\lambda < \mu$ as $P_{\rm w}(\tau)\sim \tau^{-3/2}e^{-\tau/\tau_0}$
with $\tau_0=1/(\sqrt{\mu}-\sqrt{\lambda})^2$. However, for the model with general
distributions of the number of incoming tasks, the explicit form of the
exponential correction
factor is not obvious.

In our priority queue model, the jump distance of the equivalent
random walk is unbounded to the right, whereas it is at most one to
the left.  In real queue systems, however, more than one tasks may be executed
in a unit time. Therefore, a natural extension of our model is
to allow the number of executed tasks in a unit time to exceed
one. To be specific, in addition to the heterogeneity of
the number of incoming tasks, i.e., $n$ tasks are incoming
with probability $\lambda_n\sim n^{-\gamma_{\rm in}}$ per unit time,
we can suppose that $\ell$ tasks are executed with probability
$\mu_{\ell}\sim \ell^{-\gamma_{\rm out}}$.
Our numerical results for $P_{\rm w}(\tau)$
seem to fit the formulas shown in Table~\ref{table1},
with the exponent $\gamma$ replaced by the minimum
of $\gamma_{\rm in}$ and $\gamma_{\rm out}$, as far as
both $\gamma_{\rm in}$ and $\gamma_{\rm out}$ are larger
than $2$ (not shown).
This suggests that the dominant tail determines the behavior
of the waiting-time distribution in the priority queue system.
In particular, when the distribution of the number of executed
tasks is neither binary nor heavy-tailed (e.g. purely exponential),
which may be true for many real queues, our results hold
because $\gamma=\gamma_{\rm in}$.\\

\noindent{\sf Acknowledgments}
This work was supported by the KOSEF grant for Acceleration Research (CNRC)
(Grant No. R17-2007-073-01001-0), the KRCF, and Grants-in-Aid for Scientific Research from MEXT,
Japan (Grants No. 20760258 and No. 20540382).\\

\appendix

\section*{Comparison of the Grinstein-Linsker solution and the
generating-function solution}\label{sec:GL derivation}

GL analyzed a priority queue model in which, in a
unit time, a new task arrives with probability $\lambda$ and the task with the
highest priority in the queue is executed with probability $\mu$, which
corresponds to $\lambda_0=1-\lambda$, $\lambda_1=\lambda$, and $\lambda_n=0$ for
$n \ge 2$ in our model~\cite{Grinstein}. They obtained the solution
of the waiting-time distribution by analyzing the continuous-time dynamics.
We compare the GL solution and
the solution derived via the generating function
in the GL limit.

\subsection{The queue-length distribution}

The generating function of the queue-length distribution in the steady
state is given in Eq.~\eqref{eq:calQ_x(z)} in the main text. By substituting
$\Lambda(z)=1-\lambda+\lambda z$ and
$\left<n\right>_{\lambda}=\lambda$ into Eq.~\eqref{eq:calQ_x(z)}, we obtain
\begin{equation}
\tilde{{\cal Q}}_x(z) = \frac{\mu-\lambda(1-x)}
{\lambda(\mu-1)(1-x)z+(1-\lambda+\lambda x)\mu},
\label{eq:calQ(s)}
\end{equation}
which leads to
\begin{equation}
\tilde{Q}_x(m)=\frac{\mu-\lambda(1-x)}{(1-\lambda+\lambda x)\mu}
\left(\frac{\lambda(1-\mu)(1-x)}{(1-\lambda+\lambda x)\mu}\right)^m.
\label{eq:Q(m)}
\end{equation}
Using the continuous-time approach, GL derived
\begin{equation}
\tilde{Q}_x(m)=\frac{\mu-\lambda(1-x)}{\mu}
\left(\frac{\lambda(1-x)}{\mu}\right)^m.
\label{eq:Q(m) Grinstein}
\end{equation}
Equations~\eqref{eq:Q(m)} and \eqref{eq:Q(m) Grinstein} are consistent in the
limit $\lambda, \mu\to 0$.

\subsection{The waiting-time distribution and the exponential cutoff}

To obtain the waiting-time distribution of a task,
we use the following theorem
\cite{Bender,Odlyzko,Klarner}:

\bigskip
{\bf Theorem}: Suppose that, for real numbers $s^*$ and ${\mathcal F}^*$,
a power series ${\mathcal F}(s)=\sum^{\infty}_{t=1}
a(t)s^t$ with nonnegative coefficients $a(1)$, $a(2)$, $\ldots$
satisfies the following equations~\eqref{eq:cnd1}, \eqref{eq:cnd2},
and \eqref{eq:cnd3}.
\begin{equation}
F(s,{\mathcal F}) \mbox{ is analytic near } (s,{\mathcal F})
=(s^*,{\mathcal F}^*);
\label{eq:cnd1}
\end{equation}
\begin{equation}
\mbox{if } |s|\le s^*, |{\mathcal F}|\le {\mathcal F}^*,\quad
F(s,{\mathcal F}) =
\frac{\partial F(s,{\mathcal F})}{\partial{\mathcal F}} = 0
\mbox{ if and only if } (s,{\mathcal F})=(s^*,{\mathcal F}^*);
\label{eq:cnd2}
\end{equation}
\begin{equation}
\frac{\partial F(s^*,{\mathcal F}^*)}{\partial s} \neq 0,\quad
\frac{\partial^2 F(s^*,{\mathcal F}^*)}{\partial \mathcal F^2} \neq 0.
\label{eq:cnd3}
\end{equation}

Then,
\begin{equation}
a(t)\approx \left(\frac{s^* \frac{\partial F(s^*,{\mathcal F}^*)}
{\partial s}}
{2\pi \frac{\partial^2 F(s^*,{\mathcal F}^*)}{\partial{\mathcal F}^2}}
\right)^{\frac{1}{2}}
t^{-\frac{3}{2}} s^{* -t},\quad t\to\infty.
\label{eq:theorem}
\end{equation}

\bigskip

To apply this theorem to the GL queue model, we
define
\begin{equation}
F\left(s,{\mathcal F}\right)=
s\left[\left(1-\mu\right){\mathcal F}+\mu\right]
\left\{1-\lambda+\lambda\left[\left(1-x\right){\mathcal F}+x\right]\right\}
-{\mathcal F},
\label{eq:base1}
\end{equation}
so that $F\left(s,{\mathcal F}_x\left(s\right)\right)=0$
holds, where ${\mathcal F}_x(s)=\sum^{\infty}_{t=1}
f_x(t)s^t$ is the generating function
of the first-passage time probability.
Then the other main condition of the theorem
[see Eq.~\eqref{eq:cnd2}] reads
\begin{equation}
\frac{\partial F(s,{\mathcal F})}{\partial{\mathcal F}}=
s(1-\mu)\left\{1-\lambda+\lambda\left[\left(1-x\right)
{\mathcal F}+x\right]\right\}+
s\left[\left(1-\mu\right){\mathcal F}+\mu\right]\lambda
(1-x)-1=0.
\label{eq:base2}
\end{equation}
The solution to Eqs.~\eqref{eq:base1} and \eqref{eq:base2}
with the minimum absolute values is given by
\begin{eqnarray}
s^* &=& \frac{1}{1-\lambda-\mu+2\lambda\mu+\lambda x-2\lambda\mu x+
2\sqrt{\lambda(1-\lambda+\lambda x)(1-x)\mu(1-\mu)}},
\label{eq:s*}\\
{\mathcal F}^* &=& \sqrt{\frac{\mu(1-\lambda+\lambda x)}
{\lambda(1-\mu)(1-x)}}.
\label{eq:calF*}
\end{eqnarray}

The rest of the conditions of the
theorem are satisfied with $s^*$ and ${\mathcal F}^*$
given by Eqs.~\eqref{eq:s*} and \eqref{eq:calF*}.
Equation~\eqref{eq:theorem} implies that
the tail of the first-passage time probability
decays as $f(t)\propto t^{-3/2}s^{* -t}$.
This asymptotic is also derived by directly
calculating ${\mathcal F}_x(s)\sim (s^*-s)^{1/2}$ as $s\uparrow s^*$
and using the Tauberian theorem \cite{Tauberian,Fellerbook,Weissbook}.

The generating function of the waiting-time distribution of a task is
equal to that of the queue-length distribution given by
Eq.~\eqref{eq:calQ_x(z)} with $s$ replaced by ${\mathcal F}_x(s)$.  To
calculate the asymptotic of the waiting-time distribution, we erase
$\Lambda$ by combining Eq.~\eqref{eq:calQ_x(z)} with $s$ replaced by
${\mathcal F}_x(s)$ and Eq.~\eqref{eq:calQ_x(z) 2-3}, which yields
\begin{equation}
\tilde{{\cal Q}}_x\left({\mathcal F}_x\left(s\right)\right)
= \frac{A_1 s\left[1-{\mathcal F}_x\left(s\right)\right]}
{(1-s){\mathcal F}_x(s)}.
\label{eq:calQ-F}
\end{equation}
Inserting Eq.~\eqref{eq:calQ-F} into Eq.~\eqref{eq:base1} results in
\begin{equation}
(1-\lambda+\lambda x)\mu(1-s)
\tilde{{\cal Q}}_x^2\left({\mathcal F}_x\left(s\right)\right)
+A_1\left[\left(1-\lambda+\lambda x+\mu\right)s-1
\right]\tilde{{\cal Q}}_x\left({\mathcal F}_x\left(s\right)\right)
-A_1^2 s = 0.
\label{eq:calQ*}
\end{equation}
Applying the theorem \eqref{eq:cnd1}-\eqref{eq:theorem}
with ${\mathcal F}\equiv
\tilde{{\cal Q}}_x\left({\mathcal F}_x\left(s\right)\right)$ leads to
the same equation \eqref{eq:s*}.
Therefore, the waiting-time distribution has the same asymptotic as
the first-passage time probability, that is, $P_{\rm w}(\tau)\sim
\tau^{-3/2}s^{* -\tau}$.
This asymptotic
is also derived by solving Eq.~\eqref{eq:calQ*}
as ${\mathcal P}_{\rm w}(s)\sim (s^*-s)^{1/2}$ as $s\uparrow s^*$.

To evaluate $s^*$, we denote the denominator of the RHS of
Eq.~\eqref{eq:s*} by $H(x)$, with $\lambda$ and $\mu$ fixed.
The existence of the exponential cutoff in the first-passage time and
the waiting-time
distribution is equivalent to $H(x)<1$ ($0\le\forall x\le 1$).

A straightforward calculation yields $d^2H/dx^2<0$,
$\lim_{x\uparrow 1} dH/dx = -\infty$, and that $dH/dx=0$ has a unique
solution $x=(\lambda-\mu)/\lambda$. As explained in the main text and in
previous literature \cite{Grinstein}, the analysis of case
$\lambda>\mu$ is ascribed to that of case $\lambda=\mu$.
Therefore, we assume $\lambda\le\mu$ and obtain
$dH/dx< 0$ ($0< x\le 1$). Then the maximum of $H(x)$ is realized
at $x=0$, so that the smallest $s^*$ is equal to
\begin{equation}
s^* = \frac{1}{H(0)}=\frac{1}{1-\lambda-\mu+2\lambda\mu+
2\sqrt{\lambda(1-\lambda)\mu(1-\mu)}}.
\label{eq:s* final}
\end{equation}

When $\mu=\lambda$, we obtain $s^*=1$. The asymptotic of the
waiting-time distribution is $P_{\rm w}(\tau)\sim \tau^{-3/2}$,
which is consistent with the results in \cite{Grinstein} and
coincides with our results for $\gamma>3$.

When $\mu>\lambda$, we obtain $s^*>1$ and
$P_{\rm w}(\tau)\sim \tau^{-3/2}e^{-\tau/\tau_0}$, where
$\tau_0=1/\ln s^*$. In the limit $\lambda, \mu\to 0$, our
discrete-time model tends to GL's continuous-time queue dynamics.
By inserting $\lambda=\lambda^{\prime}\Delta\tau$, $\mu=
\mu^{\prime}\Delta \tau$, and $\tau=\tau^{\prime}/\Delta\tau$
into Eq.~\eqref{eq:s* final} and letting $\Delta\tau\to 0$, we obtain
$P_{\rm w}(\tau)\sim \tau^{\prime -3/2}e^{-\tau^{\prime}/\tau_0}$,
where $\tau_0=1/(\sqrt{\mu}-\sqrt{\lambda})^2$. The predicted $\tau_0$
agrees with the one derived by GL. They concluded
$P_{\rm w}(\tau)\sim \tau^{\prime -3/2}e^{-\tau^{\prime}/\tau_0}$
for $\tau\ll \tau_0$ and $P_{\rm w}(\tau)\sim \tau^{\prime -5/2}e^{-\tau^{\prime}/\tau_0}$
for $\tau\gg \tau_0$. Our results only reproduce the asymptotic
on the intermediate timescale (i.e., $\tau\ll \tau_0$) because $\tau_0$ diverges
as $\lambda,\mu\to 0$.

%
%

\end{document}